\documentclass{vgtc}                          % final (conference style)
\ifpdf%                                % if we use pdflatex
  \pdfoutput=1\relax                   % create PDFs from pdfLaTeX
  \usepackage{graphicx}                % allow us to embed graphics files
  \DeclareGraphicsExtensions{.pdf,.png,.jpg,.jpeg} % for pdflatex we expect .pdf, .png, or .jpg files
\else%                                 % else we use pure latex
  \usepackage{graphicx}                % allow us to embed graphics files
  \DeclareGraphicsExtensions{.eps}     % for pure latex we expect eps files
\fi%

\graphicspath{{figures/}{pictures/}{images/}{./}} % where to search for the images

\usepackage{microtype}                 % use micro-typography (slightly more compact, better to read)
\PassOptionsToPackage{warn}{textcomp}  % to address font issues with \textrightarrow
\usepackage{textcomp}                  % use better special symbols
\usepackage{mathptmx}                  % use matching math font
\usepackage{times}                     % we use Times as the main font
\usepackage{cite}                      % needed to automatically sort the references
\usepackage{tabu}                      % only used for the table example
\usepackage{booktabs}                  % only used for the table example
\usepackage{diagbox}
\usepackage{multirow}
\usepackage{color}
\usepackage{soul}
\usepackage{enumitem}
\setitemize{noitemsep,topsep=0pt,parsep=0pt,partopsep=0pt}

\onlineid{1060}

\vgtccategory{Research}

\vgtcinsertpkg
\newcommand{\techname}{\textit{ConceptExplorer}}
\title{ConceptExplorer: Visual Analysis of Concept Drifts \\ in Multi-source  Time-series Data}

\author{Xumeng Wang\thanks{e-mail: wangxumeng@zju.edu.cn}\\ %
        \parbox{1.4in}{\scriptsize \centering State Key Lab of CAD\&CG\\ Zhejiang University} %
\and Wei Chen\thanks{e-mail: chenwei@cad.zju.edu.cn}\\ %
     \parbox{1.4in}{\scriptsize \centering State Key Lab of CAD\&CG\\ Zhejiang University} %
\and Jiazhi Xia\thanks{e-mail: xiajiazhi@csu.edu.cn}\\ %
     \parbox{1.4in}{\scriptsize \centering School of Computer Science and Engineering\\ Central South University}
\and{Zexian Chen}\thanks{e-mail: zexianchen@zju.edu.cn}\\ %
        \parbox{1.4in}{\scriptsize \centering State Key Lab of CAD\&CG\\ Zhejiang University} %
\and{Dongshi Xu}\thanks{e-mail: p1703085223@stu.cjlu.edu.cn}\\ %
        \parbox{1.4in}{\scriptsize \centering College of Information Engineering\\ China Jiliang University} %
\and{Xiangyang Wu}\thanks{e-mail: wuxy@hdu.edu.cn}\\ %
        \parbox{1.4in}{\scriptsize \centering Institute of Graphics and Image\\ Hangzhou Dianzi University} %
\and{Mingliang Xu}\thanks{e-mail: iexumingliang@zzu.edu.cn}\\ %
        {\scriptsize Zhengzhou University} %
\and{Tobias Schreck}\thanks{e-mail: tobias.schreck@cgv.tugraz.at \newline Wei Chen and Jiazhi Xia are corresponding authors.}\\ %
        {\scriptsize Graz University of Technology} %
}
\teaser{
   \centering
   \includegraphics[width=\columnwidth]{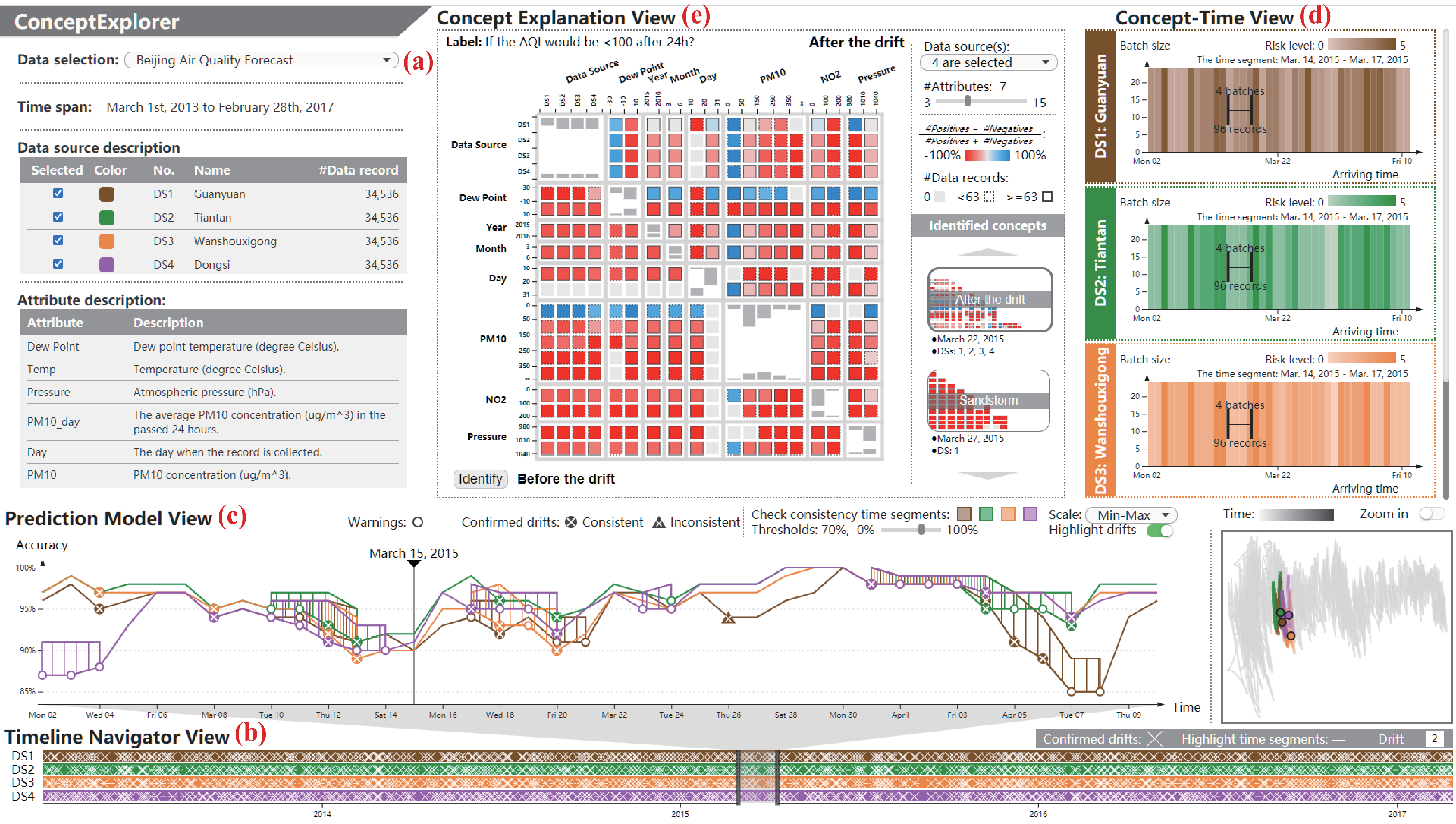}
   \caption{\techname{} contains five main views. (a) The data entrance introduces the applied data sources and attributes. (b) The timeline navigator view is used to select interested time segments; (c) The prediction model view presents the training process of prediction models to explain concept drift detection model; (d) The concept-time view shows the time segments recommended for analyzing concepts based on the moment selected by analysts in prediction model view. 
   (e) The explanation view compares concepts pairwise through a correlation matrix.}
   \label{fig:tea}
}

\abstract{
Time-series data is widely studied in various scenarios, like weather forecast, stock market, customer behavior analysis. To comprehensively learn about the dynamic environments, it is necessary to comprehend features from multiple data sources. This paper proposes a novel visual analysis approach for detecting and analyzing concept drifts from multi-sourced time-series.  We propose a visual detection scheme for discovering concept drifts from multiple sourced time-series based on prediction models. We design a drift level index to depict the dynamics, and a consistency judgment model to justify whether the concept drifts from various sources are consistent. Our integrated visual interface, \techname{}, facilitates visual exploration, extraction, understanding, and comparison of concepts and concept drifts from multi-source time-series data.  We conduct three case studies and expert interviews to verify the effectiveness of our approach.
}  % end of abstract

\CCScatlist{
  Temporal data; data analysis, reasoning, problem solving, and decision making; machine learning techniques.
}

\begin{document}


\firstsection{Introduction}

\maketitle
Facing the changing world, analysts track the underlying relationship between the interested target and the environment to understand, explain, and predict the evolving events.
We use the term \textbf{concept}~\cite{gama2014survey} to describe the underlying relationship between the interested target variable and the environment variables in time-series data.
The concepts evolve and are diverse across different data sources, \textit{e.g.}, data from different groups or regions.
We denote the change of concept as \textbf{concept drift}~\cite{schlimmer1986incremental}.
Tracking the concept drift is of theoretical and practical significance for domain experts, like portfolio selection~\cite{li2012pamr}.
It provides the knowledge of dynamical concepts, updates the understanding of underlying relationships, and sharpens the insights into the differences among various groups.

For example, financial experts are interested in the price fluctuation of stocks.
They would like to track the relationship between the stock index (\textit{i.e.}, the target variable) and various economic indicators (\textit{i.e.}, the environment variables).
By analyzing the concepts in a period, they can identify the most correlated economic indicators.
This knowledge is helpful to explain the factors and predict the trend of stock indices.
When concept drift is identified, their understanding of underlying relationships is updated with the disclosed change.
The concept drifts in different stock markets are related and, however, heterogeneous~\cite{baltussen2019indexing}.
For instance, the fluctuation of stock indices of the U.S. market would be different but related to that of the European market.
The knowledge of and comparison between the concept drifts in different markets sharpens the insights into the relationship and difference among multiples stock markets.
Another example is to inspect the cure rate during a virus spreading.
The target variable is the cure rate and the environment variables include the health profile of patients.
Knowing the concept drifts in different regions is important for experts to custom the diagnosis strategies.

Tracking concept drifts from the huge number of multi-source time-series data is technically demanding. 
There are two major challenges to be addressed.
The first one is to model concept drifts and identify them in the time-series data~\cite{keogh2004segmenting, shurkhovetskyy2018data}.
To detect certain patterns in multivariate time-series data, domain experts need to manually set the patterns and parameters. 
However, experts have not an explicit description of the underlying relationship.
The recent time-series clustering~\cite{olier2006capturing, lee2009visualization,liu2014survey} and frequency-based detection~\cite{liu2018tpflow} are usually used to detect pattern automatically without specifying the target first. 
Because they assume that the state of a process is repetitive, these approaches are suitable for periodic processes.
However, the concept drift occurrences in real-world scenarios are always evolving irregularly. 
The cluster-based or frequency-based approaches cannot achieve the desired performance.
The second challenge is to design an intuitive visualization to illustrate the evolving concept drifts over time and the relationship among concept drifts across multiple data sources~\cite{steed2017falcon}.
The concept drifts often occur hundreds of times in a rapidly changed period.
The data heterogeneity among data sources results in complicated relationships among concept drifts.
Therefore, an efficient interactive visualization is necessary to present the concept drifts over a long time span, which allows experts to focus on a selected time period in interest for further inspection.

We have developed a visual analytics system, \techname{}, to address these challenges.
To model the concept, we have employed prediction models to capture the underlying relationship between the target variable and the environment variables.
The concept drift is captured by the accuracy change of prediction models.
Based on the performance of prediction models, we have formulated the drift level index to indicate concept drifts and proposed a consistency judgment model to inspect the inconsistency among multi-sources.
To effectively convey complicated concept drifts and their relationships, we have developed an interactive visualization that combines the strengths of multiple views.
In particular, we have developed a timeline navigator view to help users quickly get an overview of occurred concept drifts in a long time span in multiple sources.
The prediction model view presents the feature of prediction models and thus sharpen the insights into concept drifts.
The concept-time view allows analysts to focus on and fine-turn the period of a certain concept.
The concept explanation view employs a matrix visualization to present the correlation between the target variable and environment variables.
It supports the comparison between two concepts by diagonally juxtaposing them in the matrix.
These visualizations are coordinated to support the interactive analysis of concept drifts from multi-source time-series data.
Lastly, we conduct three case studies with real application scenarios and collect feedback from three experts to verify the effectiveness of our approach.

The major contributions of this work are:
\begin{itemize}
  \item A visual analytics system that helps experts understand and analyze the concept drifts over time and their relationships across multiple data sources.
  \item A description of the concept drift that takes advantage of prediction models; it induces a set of models for concept drift detection and comparison.
  \item A coordinated visualization that combines a set of novel designs in the timeline visualization, matrix visualization, and prediction model view.
\end{itemize}

\section{Related Work}
We surveyed existing studies from two aspects: 1) detection of concept drifts, and 2) visualization of time-series data.

\subsection{Detection of Concept Drifts}
Concept drifts are defined as the changes in the joint distribution between the environment variables (\textit{i.e.}, the time-series data that analysts collected) and target variables(\textit{i.e.}, the labels that analysts want to predict)~\cite{schlimmer1986incremental, gama2014survey}. Approaches for detecting concept drifts are fall in two categories: performance-based approaches and distribution-based approaches. Performance-based approaches detect concept drifts from the abnormal fluctuation of performance indicators, like accuracy. Drift Detection Method (DDM)~\cite{gama2004learning} recognizes an abnormal increase in error rate over certain ranges as a warning or a concept drift occurrence. The ranges are determined by the confidence intervals of the Normal distribution. A variety of statistical test methods can be applied after the study subject transforms into the error rate.  For instance, concept drifts are identified by a set of the chi-square test~\cite{nishida2007detecting}. To avoid the influence of the size of the upcoming data, Fisher’s Exact test is chosen~\cite{de2018concept}. 

Following the definition of the concept, distribution-based approaches compare data distributions and detect concept drifts by identifying distribution changes. However, calculating distribution similarity is a time-consuming task due to the complexity of distribution characteristics, \textit{i.e.}, extreme values, skewness, variance, etc. Considering an incremental learning process, the problem can be simplified by focusing on the differences. Certain assumptions are made~\cite{dos2016fast} to describe the changes as a series of operations between constant values.  Besides, grouping variables is a prominent approach to simplify the density statistics process. The framework~\cite{sethi2016grid} maps variables into a grid space and performs density-based clustering on grid cells. Then the influences caused by the upcoming data can be summarized as the cluster generations or cluster extensions. Taking advantage of \textit{k-nearest neighbor} (KNN), sub-spaces can be constructed~\cite{liu2018accumulating} for a sample set, and density variations are identified with a distance measurement. In summary, distribution-based approaches need to be supported by intricate quantitative evaluation~\cite{lu2018learning}.%, which request expertise to users.

\subsection{Visualization of Time-series Data}

Identifying patterns from multidimensional time-series data is a comprehensive process. It is thus essential to integrate visualization and data analysis methods for better efficiency. Existing studies achieve this goal from three perspectives: setting target patterns interactively, extracting repeated patterns based on clustering and frequency features, and detect abnormal patterns by leveraging machine learning approaches.

Within a visual interface, analysts can express what they want from the visual analysis system. Thermalplot~\cite{stitz2015thermalplot} supports specify by setting weights for each attribute. Time-varying objects are mapped into a two-dimensional space defined by the degree of interest and corresponding change over time. To explore co-occurrence patterns, COPE~\cite{li2018cope} needs analysts to specify events by setting thresholds for attribute values. The spatiotemporal pattern of similar events can be checked with COPE. It is effective to allow users to gradually narrow their search, especially when they are not sure what they want. TimeNotes~\cite{walker2015timenotes} provide users with a hierarchy time axes. Users are allowed to iteratively select one or more small time range of interest from a large time range by brushing.

When analysts have limited knowledge of datasets, it is necessary to augment the analysis with automatic methods. Temporal Multidimensional Scaling (TMDS)~\cite{jackle2015temporal} discretizes the time dimension by a user-defined sliding window, and projects the multi-dimensional data in each window to one dimension via MDS. After a series of flipping operations, one-dimensional projections are juxtaposed to show temporal patterns. %By leveraging the dimensionality reduction method, the PCA can be incrementally updated and approximately maintain the position of previous points~\cite{fujiwara2019incremental}. Analysts can learn the relationship between the new arrival data record and previous ones by observing the projection of streaming data.
In addition to dimensionality reduction methods, extracting important periods can also reduce the user's workload. StreamExplorer~\cite{wu2018streamexplorer} employs a subevent detection model to identify important periods from a social stream. Because major events always lead to popular discussions, StreamExplorer recommends users the periods with a large number of tweets to analyze related events. To extract patterns flexibly, TPFlow~\cite{liu2018tpflow} employs a piecewise rank-one tensor decomposition to detect sub-tensors (\textit{i.e.} multidimensional patterns) with a top priority. %As the decomposition iterates, TPFlow can divide the spatio-temporal data into hierarchies for users to explore. 

If unpredictable events are regarded as abnormal, the high prediction error of automatic models may imply abnormal patterns~\cite{tkachev2019local}. Taking advantage of the same feature, concept drift detection can be applied to locate useful patterns from dynamic environments. Visualization techniques have been used to depict the development of concept drifts~\cite{yao2013concept, demvsar2018detecting}.
%Yao et al.~\cite{yao2013concept} recorded the concept transformation processes and demonstrated drift width (the period length of each concept drift) via a stacked bar chart from the perspective of features.
%The visual representation presented by Dem{\v{s}}ar and Bosni{\'c} includes more details, like levels of class noise~\cite{demvsar2018detecting}. Their design can be used to test the capabilities of different detection models--whether they can furnish timely and accurate warnings for distinct concept drifts.
%Consistent with the approaches to concept drift detection, two aspects were considered to further explain concept drifts.
%On the one hand, illustrating attribute distribution can benefit concept drift comprehension, as concept drift defined.
Common charts, like line charts~\cite{webb2018analyzing}, scatter plots~\cite{stiglic2011interpretability} and parallel coordinates~\cite{pratt2003visualizing} are employed for this purpose.
%To unfold the time-varying dimension,~\cite{webb2018analyzing} presented the numerical changes of a group of attributes by line charts. The same charts are also be utilized to show drift magnitudes according to marginal distribution.
On the other hand, model-generating information can convey the characteristics of concept drifts. The time-varying contributions of each attribute value can be visualized for classification~\cite{demvsar2014visualization}. By marking the fluctuations, the occurrences of concept drifts can be easily identified from a micro-level. Also, users are allowed to take an overview from a macro level to assess the importance of each attribute value~\cite{demvsar2014visualization}.
None of these studies can explain why a concept drift is identified by the detection model, which is viral for the final decision.

\section{Problem Definition and Models}
Two models are applied to support our goals. Before introducing the goals and the models, we first explain related definitions.
\subsection{Definitions}
\label{sec:def}
In this work, time-series from multiple sources are presented as temporal \textbf{data records}.
Each data record contains multiple \textbf{environment variables} $X$ and a \textbf{target variable} $y$.
The data records are distributed non-uniformly along the timeline.
We group data records in a unit time segment into a \textbf{batch}.
A batch forms a basic unit for coordinated analysis among different data sources. 
We denote the data source and timestamp of a batch as its \textbf{context}.

In machine learning scenarios, records are usually used by the training model, where the target variable is the data label.
Without loss of generality, the target variable is limited to be \textit{binary} in this paper.
It is assumed that there is an underlying relationship between the environment variables and the target variable, e.g., an underlying mapping $y = f(X)$, or a conditional distribution $p(y\mid X)$~\cite{gama2014survey}.
A \textbf{concept} refers to such a relationship.
The changes of concept over time is denoted as \textbf{concept drift}~\cite{widmer1996learning}.
%\subsection{The Online Learning Model}
%For each data source, the online learning scheme~\cite{ditzler2015learning} is employed to learn the analyst-specified concept. The input and output of the online learning model are attribute values of time-series, and a set of binary labels, respectively. As shown in Figure~\ref{fig:olm} (a), the label arrives after the input is collected. The online learning model first predicts the label according to the input. The pair of time-series and associated label is regarded as the training data, and triggers an iteration to update the weights of the online learning model. The prediction accuracy (or error rate) is updated after comparing the result of the previous prediction with the verified labels.

% \begin{figure} [!htbp]
%    \centering
%    \includegraphics[width=0.99\columnwidth]{pictures/olm.eps}
%    \caption{Explanation of the online learning model trained with time series data. }
%    \label{fig:olm}
% \end{figure}

\subsection{Goals}
\label{sec:goa}
Our main goal is a visual analytics tool for identifying and understanding concept drifts in multiple time-series data.
We identify four goals in building such a visual analytics system:

\textbf{G1: Automatic identification of concept drifts and concepts.} It is laborious to browse the time-series data through the entire time span. Moreover, the concept and concept drift are only implicitly embedded in the time-series data. Identifying numerous concepts and concept drifts individually is cumbersome. Therefore, the first goal of our system is to support the automatic identification of stable concept and important concept drifts.

\textbf{G2: Visual representation of concepts.} There is not an explicit definition of concepts. The assumed definitions, including the mapping or the conditional probability, are complicated to describe and understand. Therefore, presenting visualization to disclose the pattern of concepts, \textit{i.e.}, the relationship between the target variable and environment variables is needed.

\textbf{G3: Discrimination of concept drifts in multiple data sources.} With the same target variable and environment variables are collected, concepts drifts could be heterogeneous in different data sources~\cite{hochman2012visualizing}. When an inconsistency occurs among different data sources, analysts should be able to discriminate them and verify the interested concept drift.
%\item{\textbf{G1: Provide quantitative description of drift status.} It is laborious to browse the concept during the entire time span. When the analysts has comprehend the existing concept, it is meaningless to continue analyzing the corresponding data. Seeking a efficient analysis process, analysts need a guide to locate the significant intervals, which can benefit from quantitative descriptions~\cite{cao2017voila,zhang2018idmvis}. Quantified indicators, i.e., drift level, can provide better guides than binary description, like a concept drift is occurred or not.}
%\item{\textbf{G2: Compare concept drift occurrences in different data sources.}  Although there exist differences in time or intensity, the data collected from the same environment may be influenced similarly. Thus, concept drift occurrences in different data sources could be identical~\cite{santillana2015combining}. Analysts can determine if a data source is consistent with the overall environment according to the status of other data sources. If an exception occurs, analysts need to emphasize on the inconsistent data sources.}
%\item{\textbf{G3: Purify concepts for further exploration.} Different concepts can be derived from the data records with different contexts~\cite{hochman2012visualizing}. It is challenging to explore the mixed information from various concepts, which may interrupt each other. Because the superimposed patterns may be blurred, effort should be made to purify the analyzed concept.}

\textbf{G4: Interactive specification of concepts.} Concepts extracted from the batches with different contexts may be numerous. Although automatic models can augment the selection process, there is a natural need for interactive exploration and specifications of concepts~\cite{lu2017state, endert2017state}.

\subsection{The Drift Level Index}
\label{sec:det}
Conventional machine learning approaches provide a quantitative description of concept drifts (\textbf{G1}).
The basic idea is that if the concept is stationary over time, the performance of a trained prediction model should be stable or increasing.
Otherwise, the prediction accuracy will decrease and trigger an index of concept drifts.
Therefore, the decrease of the prediction accuracy is a meaningful index for concept drifts.

When a concept drift occurs, the pre-trained prediction model might have a decreasing performance and be less sensitive to subsequent concept drifts.
Therefore,  the prediction model should update with the evolving of concepts, and hence the learning process is iterative, \textit{i.e.}, learning parameters for each attribute are updated when a new data record comes. Following the technique presented in~\cite{brzezinski2014combining}, we maintain a set of prediction models and take the one with the highest accuracy in the last verification as the output model. The weakest model in the set is replaced with a newly trained model. Models trained on the new data can learn new characteristics of the label and be adaptable to dynamic environments.
As a result, the prediction process has a high performance and is sensitive to the upcoming concept drifts.

\begin{figure} [!htbp]
   \centering
   \includegraphics[width=0.99\columnwidth]{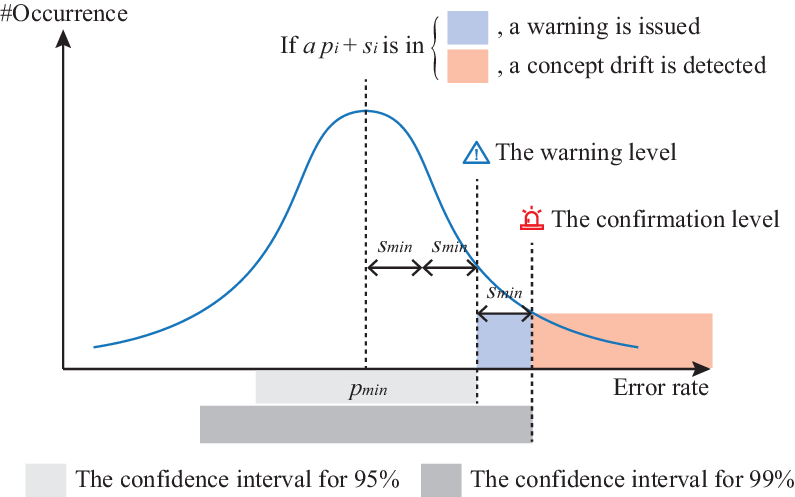}
   \caption{Explanation of the drift detection model~\cite{gama2004learning}. }
   \label{fig:ddm}
\end{figure}

%Based on the updating prediction model, we propose a quantitative index for concept drifts.
We use $p_i$ to denote the error rate of the prediction model in a sliding window, which ends at the $i$th record.
The sliding window is set to cover $500$ data records in our implementation. The distribution of correct predictions in a sliding window can be regarded as a binomial distribution. Therefore, for each $p_i$, we compute the standard deviation as $s_i = \sqrt{p_i(1-p_i)/n}$. Similar to the work in~\cite{gama2004learning}, we assume that the error rates are distributed normally. The concept drift level can be measured by the confidence levels of corresponding confidence intervals. As discussed above, in a static environment, the error rate of a prediction model is supposed to be decreasing or approximately stable over time. Although the fluctuation of the error rate is normal, the degree of its increasing typically indicates a high probability of the occurrence of a concept drift. As shown in Figure~\ref{fig:ddm}, the probability of a concept drift is computed with the minimum of the error rates $p_{min}$ (after the latest concept drift) and the standard deviation $s_{min}$~\cite{gama2004learning}. With the verified record $i$, the drift level $r_i$ is defined as:
\begin{equation}
r_i = \frac{p_i + s_i - p_{min}}{s_{min}}
\end{equation}
The threshold to determine whether a concept drift appears, \textit{i.e.}, the confirmation level, is set to be $r_i \geq 3$, which implies that the confidence level of a concept drift occurrence exceeds $99\%$~\cite{gama2004learning}. Also, when  $r_i \geq 2$ (\textit{i.e.}, the warning level, the corresponding confidence level is over $95\%$), a warning is issued. The drift level $r^t$ for a batch is regarded as the average of $r_i, r_{i+1}, ..., r_j$, where $i, i+1, ..., j$ denote the output sequence from the data records in the batch.

\subsection{The Consistency Judgment Model}
The similarity between concepts derived from different data sources can be represented by the parameter similarity of the prediction models. However, parameters of the prediction models can not summarize the consistency of concept drifts. To support \textbf{G3}, we need to learn about the response of each data source to the dynamic environment. The dynamic environment can be described by the time-varying drift levels of data sources, which are continuously captured with the training of prediction models. Supposed that a data source has a consistent response with others, the Na\''ive Bayes theory is employed to infer the time segment of the concept drift occurrence. The judgment about whether the detected concept drifts satisfy the inferred results can be concluded.

Considering the differences among data sources, we propose a consistency judgment model that is used for each data source separately. At time $t$, the drift levels of $m$ data sources $d_1,d_2,...,d_i, ...,d_m$ are recorded as $r_1^t, r_2^t,...,r_i^t, ...,r_m^t$.  Regarding $x^t_i =[r_1^t,...,r_{i - 1}^t, r_{i + 1}^t, ...,r_m^t] $ as inputs of the judgment model, the corresponding label $y_i^t$ of data source $d_i$, is defined as whether a concept drift occurs in the recent time segment $\Delta t$ (normally as a unit time segment), that is, whether the confirmation level exceeds by one of $r_i^{t - \Delta t}, r_i^{t - \Delta t + 1},..., r_i^{t+\Delta t}$. Based on the Na\"ive Bayes theory, a probability curve can be generated to depict the probabilities of concept drifts over time. The curve segments, where the probability is higher than a user-defined probability threshold $c$, implies the corresponding labels are judged as ``yes''. On the contrary, the labels of the rest time segments are ``no''. Therefore, the time segment $[t_{start}, t_{end}]$ for an occurrence of a concept drift is inferred. If the label $y_i$ is independent with $x_i$, the corresponding probability curve would be flat and has less chance to be higher than the threshold. That is, if the data source $i$ is inconsistent, the consistency judgment model results in zero time segment.

We compare the time points of detected concept drifts with the time segments to verify if a data source drifts in an inconsistent way. When the label is defined, the corresponding concept drifts should appear in the time segment $[t_{start}-\Delta t, t_{end}+\Delta t]$. A data source is regarded as inconsistent with the entire environment at $t$ if its previous concept drift occurs out of the previous time segment. Thus, the larger the parameter $c$ is, the higher the probability that data sources are considered as inconsistent.

\section{System Overview}
\subsection{Design Requirements}
To address the goals mentioned in Section~\ref{sec:goa}, five design requirements are identified to guide system design.

\textbf{DR1: Provide an overview of concept drift occurrences over time.} The drift occurrences over time can help analysts identify the interesting time segment. When the analysts' preferences are unknown, interactive and hierarchical exploration of occurrences along time intervals is needed~\cite{niederer2017taco, walker2015timenotes}.%selecting time segments need to be navigated~\cite{niederer2017taco}\hl{, sometimes multi-hierarchy}~\cite{walker2015timenotes}.

\textbf{DR2: Integrate features of concept drifts from the prediction models.} Concept drifts hinder existing prediction models (the model trained from historical data) from accurate predictions in new environments. The accuracy fluctuation of the prediction model indicates the occurrence of concept drifts~\cite{stiglic2011interpretability, becker2007real, yang2020diagnosing,liu2018analyzing, yuan2021survey}. In addition, parameters of prediction models can reflect the relationship between different inputs and the label, that is, the model's understanding of the concept~\cite{cassidy2014calculating}. 

\textbf{DR3: Identify the context of concepts and allow adjustments.} Analysts need to know the context of the analyzed data records. Considering that analysts may miss details or may disagree with the navigation, interactive adjustments for recommended results are needed~\cite{law2016vismatchmaker, liang2017photorecomposer}.

\textbf{DR4: Study the relationship between attributes and labels.} While concepts have not an explicit definition, it is essential to provide a visual explanation. The relationship between labels and attributes are considered to be an important description of concepts~\cite{gama2014survey,ditzler2015learning}.

\textbf{DR5: Compare concepts in different contexts.} Comparing different concepts facilitates the understanding of the evolving concepts and their contexts, \textit{e.g.}, the trends and outliers. There is a need to compare a newly identified concept with previously studied ones and record the identified concepts~\cite{webb2018analyzing}. Comparing concepts related to a concept drift also favors the understanding of the drift.

\subsection{Workflow}
\label{sec:wor}
\begin{figure} [!htbp]
   \centering
   \includegraphics[width=0.99\columnwidth]{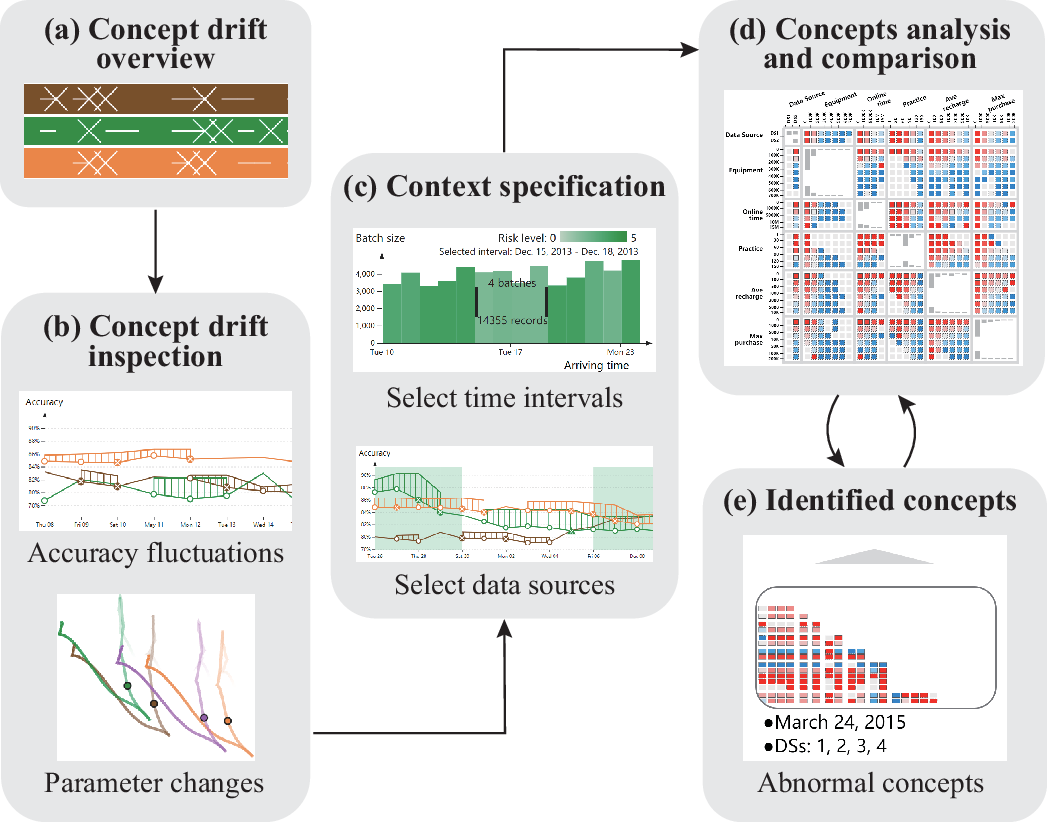}
   \caption{The five-step workflow: (a) Observing the distribution of concept drifts and warnings; (b) Inspecting concept drifts through accuracy fluctuation and parameter changes; (c) Specifying the context of the concept to be analyzed; (d) Analyzing and comparing concepts based on correlations; (e) Identifying interesting concepts. }
   \label{fig:ppl}
\end{figure}

To derive concepts from multi-source time-series datasets, analysts need to manipulate the data and select contexts. %Following the mantra of Shneiderman--``Overview First, Zoom and Filter, Then Details-on-Demand''~\cite{shneiderman1996eyes}, w
We design a five-step workflow, as shown in Figure~\ref{fig:ppl}.

We provide an overview of all concept drifts detected from all data sources during the entire time span (see Figure~\ref{fig:ppl}(a), \textbf{DR1}). Analysts may be attracted by time segments when a single data source has interesting patterns (\textit{e.g.} dense occurrences or a periodicity), or multiple data sources need to be compared (\textit{e.g.} abnormal concept drifts). When a time segment is selected, concept drifts from the prediction models are shown (\textbf{DR2}). As shown in Figure~\ref{fig:ppl}(b), the accuracy fluctuation and parameters of prediction models can assist the detection of concept drifts. Next, analysts can specify the context of the concept to be analyzed with the external knowledge of concept drifts, as shown in Figure~\ref{fig:ppl}(c). \techname{} can recommend the time segment between two adjacent concept drifts according to an analyst-specified time point. The recommended time segments for different data sources may be different, or even inconsistent. \techname{} assesses the consistency of data sources by the consistency judgment model and recommends the group of data sources with consistent concept drifts. The recommended selection is displayed (\textbf{DR3}). If analysts are not satisfied with the recommendations, they can make flexible adjustments on contexts to support special analysis tasks. To explore the concepts with specified contexts, the relationship between attributes and concepts (\textbf{DR4}, see Figure~\ref{fig:ppl}(d)) are visualized. Analysts can identify and record significant concepts that may be involved in subsequent analysis. The identified concepts can be compared with other concepts (\textbf{DR5}, see Figure~\ref{fig:ppl}(e)).

\section{ConceptExplorer}
As shown in Figure~\ref{fig:tea}, \techname{} consists of a data entrance (see Figure~\ref{fig:tea}(a)) and four views. The data entrance lists the label definition, description of data sources, and attributes. The online system is available through the link: \url{http://101.132.126.253/}.

%\textbf{DG3: Inspect the prediction accuracy of online learning models.} Improving the accuracy of model prediction is the ultimate goal of detecting concept drift and adjusting the model. It is indispensable to allow analysts to learn about the prediction accuracy. Besides, the accuracy of prediction results can indicate concept drifts intuitively. Classic concept drift detection models, like DDM~\cite{gama2004learning}, employ accuracy to determine if a concept drift has occurred. When the accuracy drops abnormally, different criteria~\cite{gama2004learning, sidhu2018novel, ren2018knowledge, pesaranghader2016fast} are proposed to judge whether concept drifts are occurred. To avoid analysts disagree with the inference from the automatic model, analysts should be allowed to learn about the evidence, i.e., the fluctuation of prediction accuracy.
\subsection{The Timeline Navigator View}
As required by \textbf{DR1}, the timeline navigator view presents the entire timeline and the indices of concept drifts from multiple data sources (see Figure~\ref{fig:tea}(b)). Each row corresponds to a data source. \techname{} assigns a unique color to each data source. Due to the limited horizontal space, the distribution of concept drifts may be dense. \techname{} employs a ``$\times$'' to mark a concept drift, which can highlight the specific moment by its intersection. Time segments, in which the drift level exceeds a certain value (initialized as the warning level, namely, $2$) are highlighted by ``$-$''. These marks indicate various patterns along the timeline, like dense occurrences, outliers, inconsistency with other data sources, periodicity.
\subsection{The Prediction Model View}
The prediction model view (Figure~\ref{fig:tea}(c)) supports \textbf{DR2}. 
\subsubsection{The Accuracy Fluctuation Chart}
\begin{figure} [!htbp]
   \centering
   \includegraphics[width=0.99\columnwidth]{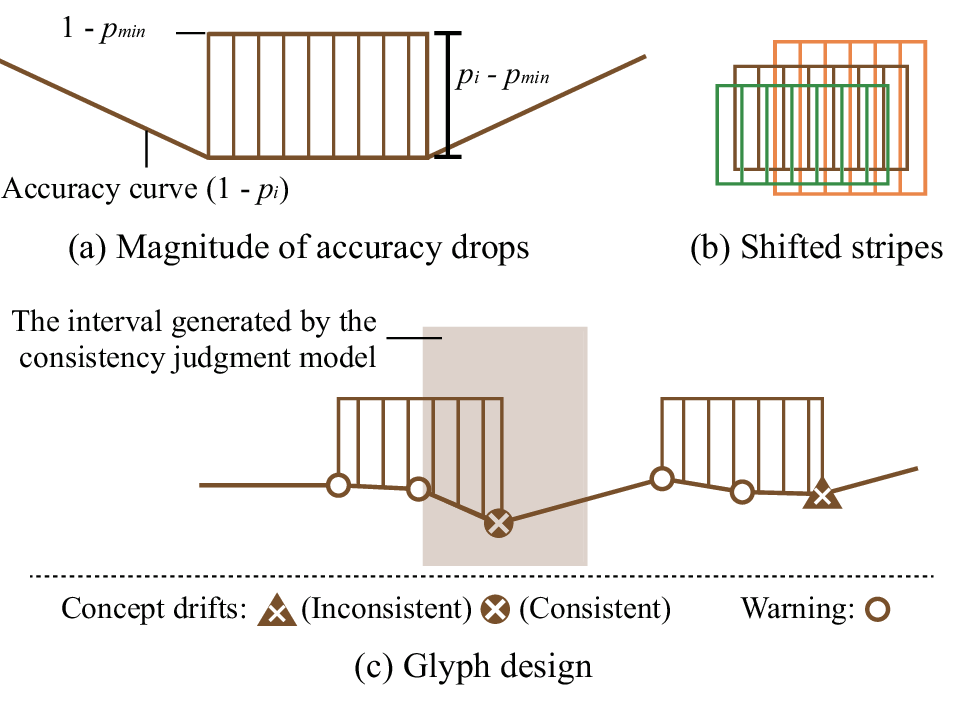}
   \caption{The visual designs for explaining the detection model and the consistency judgment model. (a) The explanation corresponds to the formula mentioned in Section~\ref{sec:det}. (b) Strips are shifted to avoid overlaps. (c) Encodings of concept drifts and warnings.}
   \label{fig:str}
\end{figure}

The line charts on the left (Figure~\ref{fig:tea}(c)) show the accuracy fluctuation of the prediction models trained by the data from each data source. The occurrences of concept drifts are labeled by ``$\times$'', which is the same as that in the timeline navigator view. In addition, the moments with warnings are encoded by hollow dots. To explain concept drift detection, the accuracy fluctuation chart visualizes the magnitude of the accuracy drop of the time segments whose drift levels are above the warning level (Figure~\ref{fig:str}(a)). Different data sources may issue drift warnings at similar time segments. To avoid misunderstandings caused by overlaps, shifted stripes are employed to highlight warning time segments (Figure~\ref{fig:str}(b)). It can be seen that even when the warning segments of different data sources are staggered, the start and end moments of different time segments can be clearly distinguished. Vertical stripes are used because they can emphasize the height, that is, the magnitude of accuracy drops. The results from the consistency judgment model are also shown in the accuracy fluctuation chart. If a concept drift is detected during a time segment that is not included by the result from the consistency judgment model, we emphasize them by a triangle mark to distinguish from circles representing others (Figure~\ref{fig:str}(c)).

\subsubsection{The Projected Parameter View}
\begin{figure} [!htbp]
   \centering
   \includegraphics[width=0.99\columnwidth]{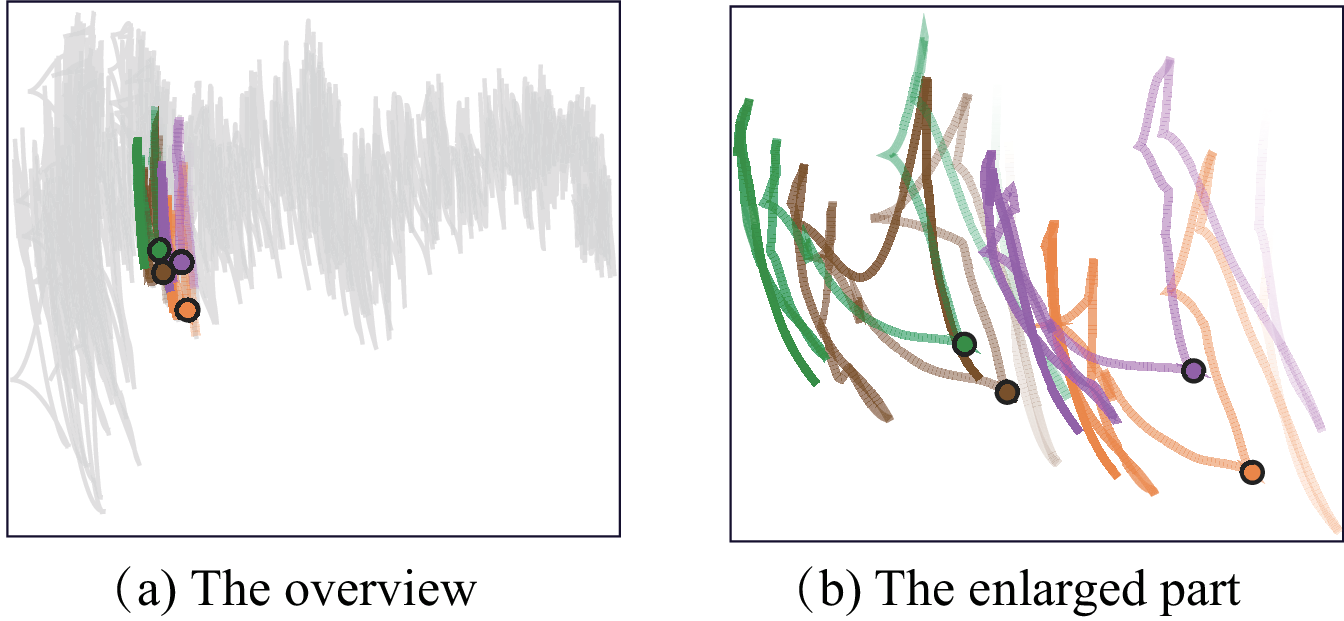}
   \caption{The parameter projections of prediction models running for all data sources. The opacity encodes the time order. (a) The overview of the entire time range. (b) An enlarged part.}
   \label{fig:zoom}
\end{figure}

The model parameters updated after each batch during the entire training process are projected into a two-dimensional plane using principal components analysis (PCA) based on singular value decomposition (SVD)~\cite{wall2003singular}. The points projected by parameters of the same data source are connected in order to form a curve. The distance between each pair of projected points illustrates the similarity of corresponding model parameters, namely, the concept similarity described by prediction models. Evolution patterns, like intersections and bundles~\cite{bach2016time, hinterreiter2020exploring} can be identified from the formed curves~\cite{ceneda2016characterizing}. The convergence and dispersal of curve segments representing different data sources indicate the agreement and disagreement of related models on the understanding of the concept. The curve segments corresponding to time segments selected in the timeline navigator view is colored by transparency, from which analysts can learn about the temporal order. The view is automatically zoomed in or out so as to fit the curves of the entire training process or the selected time segment within the window, as shown in Figure~\ref{fig:zoom}. With the background of the entire trajectories (Figure~\ref{fig:zoom}(a)), analysts can better measure relative distances. After zooming in, it can be seen (Figure~\ref{fig:zoom}(b)) that curves are not overlapped but with similar directions, that is, data sources have similar drifts. Analysts can drag the handle on the time axis of the accuracy fluctuation chart to move the circles, which highlight the projected parameters corresponding to the same moment.

\subsection{The Concept-Time View}
The concept-time view displays the time segments in different data sources that are integrated for concept analysis, as shown in Figure~\ref{fig:tea}(d). To display the sources of the applied data records, as mentioned in \textbf{DR3}, each data source is listed in a row to distinguish different data sources. Then, their data records are divided individually to introduce a specific time. Analysts need to make the trade-off between the number of data records and the clarity of concepts for appropriate adjustments. To facilitate decision-making, the drift level and the size of each batch are encoded with the color and height of the bar, respectively.  The batches that compose the data records to be analyzed are highlighted. The number of these batches and the total number of the related data records are counted. 

%\subsection{Attribute Correlation View}
%To satisfy \textbf{DR4}, the correlations between top five related attributes and the concept are summarized in the attribute correlation view, as shown in Figure~\ref{fig:tea} (d). To preserve the characteristics of data records with different contexts, data sources are separated, and the correlation for each batch of data records is quantified by the cosine similarity.  The correlations of each attribute are described by the cumulative distribution. The larger the area enclosed by the cumulative distribution curve and the two coordinate axes, the higher the correlation between the attribute and the concept is. \techname employs juxtaposed charts to facilitate visual comparison.

\subsection{The Concept Explanation View}
A correlation matrix (Figure~\ref{fig:tea}(e)) is employed to support \textbf{DR4} because of its representation ability~\cite{suschnigg2020exploration, wang2017utility}. The data source is considered as an attribute to label the context of the data records. For other attributes, the correlation for each batch of data records is quantified by the cosine similarity. The attributes are sorted by the average correlation of the selected batches.
\techname{} draws correlation matrices for the data source and analyst-specified number of the attributes with the highest correlations subject to the concept. 

For each cell, the horizontal and vertical axes of each matrix are defined by two attributes. A square in non-diagonal cells represents a set of data records whose two attributes fall into the value ranges which are encoded with the position of the square. The differences between the number of records with positive labels and those with negative labels are counted for each square. The ratio of the difference of two counts over their sum (\textit{i.e.}, $\frac{\#Positives - \#Negatives}{\#Positives + \#Negatives}$) is encoded in color (ranging from red to blue). When the label distribution in the dataset is nonuniform, analysts can reset the color mapping and encode the percentage difference in all data records in white. In some specific contexts, certain cells may be empty. To distinguish squares without a record, strokes are added in the squares with more than one record. Darker strokes imply that the record number of the square is larger than $5\%$ of the amount of chosen data records. The matrix view exhibits a symmetrical layout. Taking advantage of this feature, the current correlation pattern can be compared with the other one. Each cell on the diagonal presents a pair of histograms (i.e., a grounded histogram for lower-left corner and an inverted histogram for upper-right corner).

\subsection{Interactions}
Following the workflow mentioned in Section~\ref{sec:wor}, \techname{} supports the following interactions.

\textbf{Navigate by overview.} Analysts first brush a time segment in the timeline navigator view and check related details in the prediction model view and the concept-time view. 

\textbf{Inspect concept drifts.} In the accuracy fluctuation chart, the probability threshold $c$ for the consistency judgment model can be defined by a slider. To study why a concept drift is emphasized, analysts can check the inferred time segments for a data source. 

\textbf{Specify the context of a concept.} Analysts can specify a timestamp by dragging the handle in the accuracy fluctuation chart. According to the time stamp, the concept-time view shows the recommended time segments and selects the batches in the time segments. If analysts are unsatisfied with the automatically selected batches, they can adjust the selection ranges by dragging the boundaries. Analysts can choose the data records which need to be included in the concept explanation view. Data sources are labeled with ``inconsistent'' and ``consistent''. The set of all consistent data sources is recommended. 

\textbf{Identify concepts.} The data records with the specified context are integrated into the correlation matrix. Analysts are allowed to set the number of listed attributes. If analysts are interested in the pattern shown in the lower-left corner of the correlation matrix, i.e., the description of the concept with the current selected context, they can save the screenshot and related concept (see the right of Figure~\ref{fig:tea}(e)) by clicking the ``Identify'' button. 

\textbf{Compare concepts.} In subsequent explorations, analysts can change the data in the upper-right corner of the correlation matrix. \techname{} highlights a pair of squares at symmetrical positions for comparison when one of them is specified by analysts.

\section{Case Studies}
\label{sec:cas}
We present three case studies based on real-world datasets. Various concepts and concept drifts are analyzed to evaluate the effectiveness of \techname.
\subsection{Beijing Air Quality Forecast}
In this case, we attempt to understand the dominant factors affecting air quality. We employ the air pollutant data~\cite{zhang2017cautionary} from four nationally-controlled air-quality monitoring sites in Beijing, which was collected every hour from March 1st, 2013 to February 28th, 2017 ($34,536$ data records per site). $22$ meteorology-related dimensions are applied to predict if the air quality index (AQI) is higher than $100$ (\textit{i.e.}, worse than mild pollution) after $24$ hours.

\begin{figure} [!htbp]
   \centering
   \includegraphics[width=0.99\columnwidth]{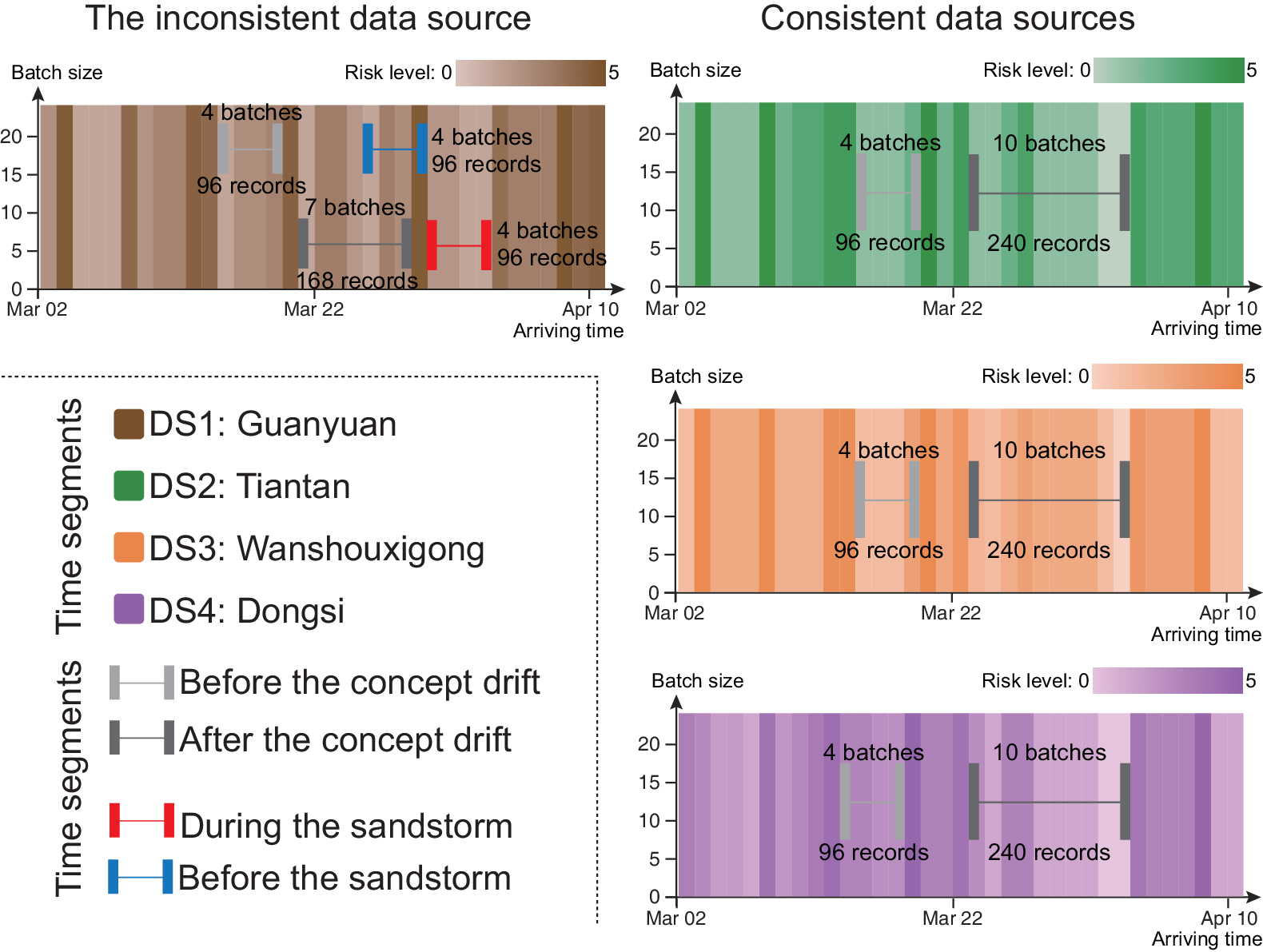}
   \caption{Details of three time segments analyzed in the first case.}
   \label{fig:air_dd}
\end{figure}

The timeline navigator view indicates that concept drifts occurred almost every few days (Figure~\ref{fig:tea}(b), \textbf{DR1}). The drift levels of all data sources are abnormally stable ($<2$, \textit{i.e.}, the warning level) during the week at the end of March 2015, except for data source (DS$1$, \textit{i.e.}, the site at Guanyuan park). We choose a $40$-day time segment around the week. As shown in Figure~\ref{fig:tea}(c), all data sources have a similar fluctuation of the prediction accuracy (\textbf{DR2}). Each one experienced more than one concept drift between March 17th and March 20th. We select two time segments before and after the drift time segment (see the black and blue time segments in Figure~\ref{fig:air_dd}, \textbf{DR3}). 

As shown in Figure~\ref{fig:tea}(e), the comparison result of two time segments indicates that their associated concepts have similarities (\textbf{DR5}). For instance, the higher the \textbf{PM10} concentration is, the more records are labeled with poor air quality (\textbf{DR4}). The difference mainly lies in that more high air quality records are observed (\textit{i.e.}, more blue squares in the upper-right corner) after the drift time segment. The records with a low \textbf{Dew Point} (\textit{i.e.}, dew point temperature ($^\circ$C)) are more likely to be labeled with good air quality. Also, the order of the attributes indicates that the dominant pollutant \textbf{PM10} is replaced by \textbf{PM2.5\_day} (\textit{i.e.}, the average PM2.5 concentration ($\mu g/m^3$) in the past $24$ hours) after the concept drift.

%The attribute correlation view indicated that the attribute with the most contribution to the prediction is \textbf{PM10\_day}. As shown in the lower-left corner of Figure~\ref{fig:tea} (e), regardless of the data source, the higher \textbf{PM10\_day}, the more records are labeled with poor air quality. Besides, we found clear patterns in the fifth row of the correlation matrix, which show the correlation of the attribute \textbf{day} (date). We review the weather records and found that there was a sandstorm in Beijing at the end of March. The direct impact of sandstorms is the significant rise of $PM10$. As the sandstorm ended, the air quality improved at the beginning of April. To further purify the sandstorm concept, we attempt to adjust the recommended interval to focus on the the end of March. As shown in Figure~\ref{fig:air_dd} (b), the drift levels in the former part and the latter parts of the recommended intervals actually reach the warning level, which implies the upcoming concept drifts. After removing the latter parts, the entire correlation matrix turned into red, as shown in the upper-right corner of Figure~\ref{fig:tea} (e). Because the air quality during the shortened intervals has remained poor. Consistent labels make the prediction task simple for online learning models. This is the reason that the prediction accuracy is extremely high.
%To our surprise, the prediction accuracy of the online learning models shown in the online learning viewer has risen to $100\%$ in a couple of days, as shown in Figure~\ref{fig:tea} (b). 

\begin{figure} [!htbp]
   \centering
   \includegraphics[width=0.7\columnwidth]{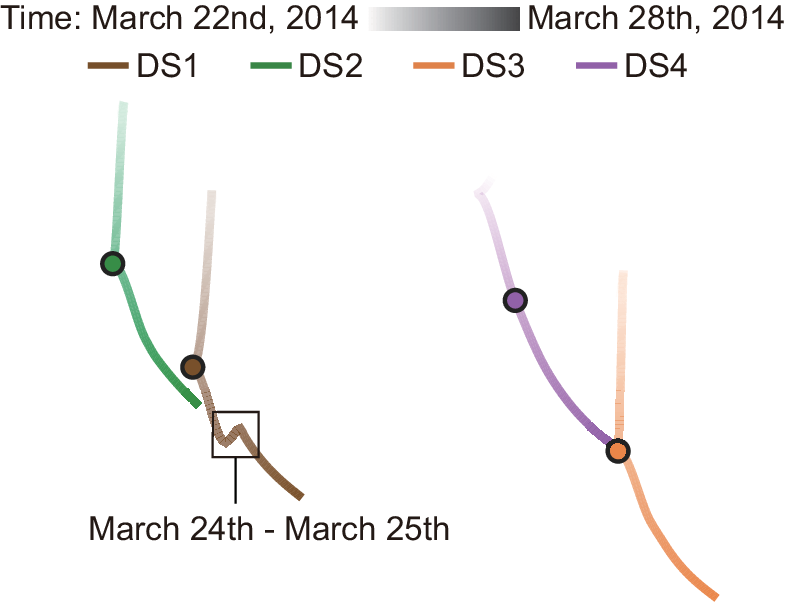}
   \caption{The parameter projection view between March 22nd, 2014 to March 28th, 2014.}
    \label{fig:air_wpv}
  \end{figure}

\begin{figure} [!htbp]
   \centering
   \includegraphics[width=0.9\columnwidth]{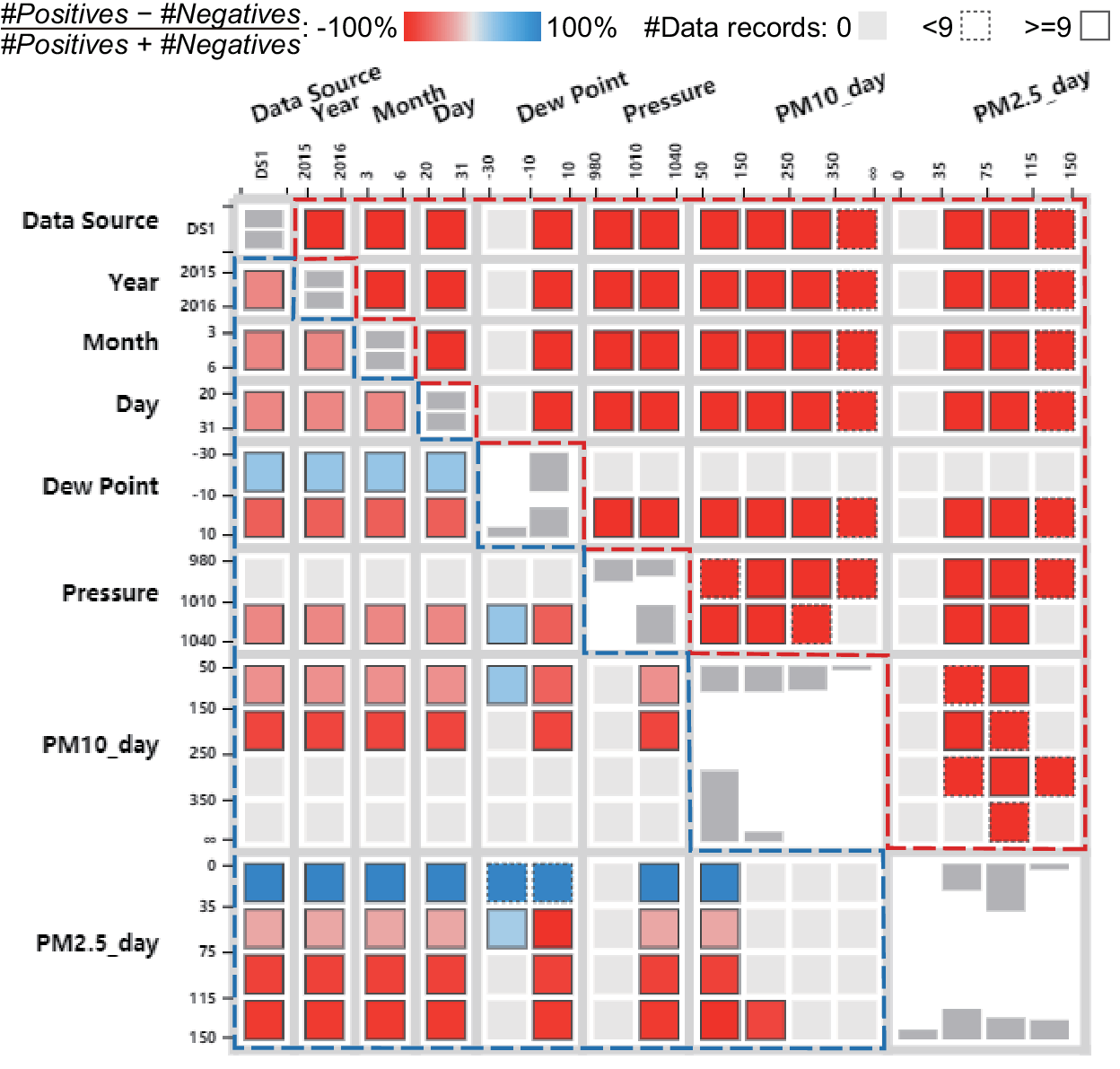}
   \caption{The correlation matrix compares the two concepts from two time segments of DS1. The lower-left corner (the dashed region in blue) corresponds to the blue time segment in Figure~\ref{fig:air_dd}. The upper-right corner (the dashed region in red) exhibits the sandstorm pattern.}
    \label{fig:air_ma}
\end{figure}

The concept drift of DS$1$ occurred on March 26th is identified as inconsistent with others by the consistency judgment model. Besides, the projected parameter trajectory of DS$1$ (see Figure~\ref{fig:air_wpv}) indicates that the parameters of DS$1$ go through a twist that is different from others (\textbf{DR2}). To explore the inconsistent behavior of DS1, the time segments after (the red time segments) the concept drift on March 26th for DS1 is checked (Figure~\ref{fig:air_dd}, \textbf{DR3}) . All cells in the correlation matrix turn into red (see the upper-right corner highlighted by the red dashed line in Figure~\ref{fig:air_ma}), which implies that almost all records are labeled with poor air quality (\textbf{DR4, DR5}). The weather records indicate that there was a sandstorm in Beijing at the end of March. The dominant pollutant during the time segment before the sandstorm (see the blue time segments in Figure~\ref{fig:air_dd}) is PM10, as shown in the lower-left corner highlighted by the blue dashed line of Figure~\ref{fig:air_ma}, which contributes to the inconsistent concept drift. 

%mainly dominated pollutant of air quality in the previous interval was \textbf{PM2.5\_day} (i.e., the average $PM2.5$ concentration ($\mu g/m^3$) in the past $24$ hours), which is replaced by \textbf{PM10\_day} (i.e., the average $PM10$ concentration ($\mu g/m^3$) in the latter interval.

Actually, other data sources record the same sandstorm. The dominance of PM2.5 has not been replaced by PM10 before the sandstorm, and thus the concept drift was not triggered by the sandstorm. Instead, the same label with poor air quality make the prediction task simple for the prediction model. The prediction accuracy has risen to $100\%$ in a couple of days. At the beginning of April, the sandstorm ended, and the air quality detected by all data sources improves, which leads to the next concept drift. An expert working in the meteorological bureau told us that the spring sandstorms in Beijing are basically caused by PM2.5. Such pollutants are spread by wind. Therefore, the detection results of sites in different locations have slight differences.

\subsection{Consumption Behaviors of MMORPG Players}
In the second case, we study the dynamics of consumption behaviors in multiplayer online role-playing game (MMORPG) to understand game company's operating strategies. For example, releasing a new role may attract new players to join in the game and consume, which leads to changes in the concept of consumption behaviors. The employed dataset contains player records from three servers ($647,800$ player records from Server17, $702,125$ player records from Server164, and $585,048$ player records from Server230) of a MMORPG from August 16th, 2013 to January 19th, 2014. Three servers were started at different timestamps: Server17, Sever164, and Server230, which are in order of time, that is, players on different servers register for the game at different time periods. For each player, $21$ attributes, like \textbf{equipment} (\textit{i.e.}, the combat effectiveness score of the player's equipment), \textbf{practice} (\textit{i.e.}, the level of practice, improved by learning and improving skills and finishing tasks), are recorded every day. The consumption records for the upcoming week of players form a group of time-series. %Existing operating models are expected to be learned from relationships between consumption records and other behaviors. %Studying these behaviors favors the maintaining of the game. 

\begin{figure} [!ht]
   \centering
   \includegraphics[width=0.8\columnwidth]{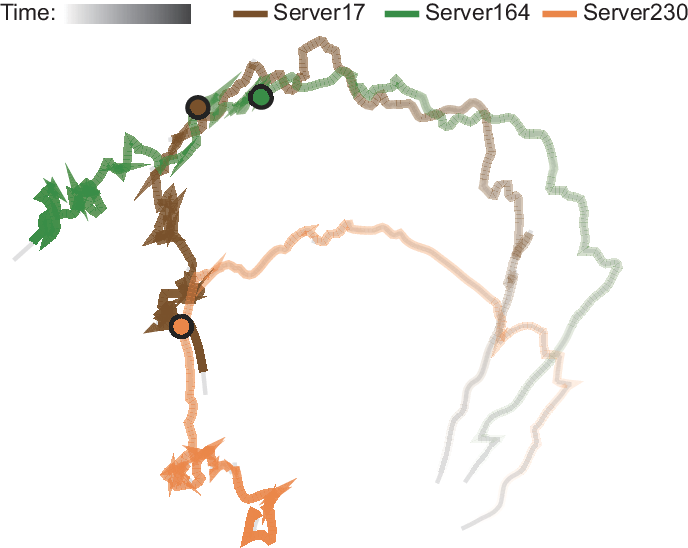}
   \caption{The overview of the parameter projection view. The orange circle denotes the moment when Server230 is merged. }
   \label{fig:net_wpv}
\end{figure}

\begin{figure} [!ht]
   \centering
   \includegraphics[width=0.9\columnwidth]{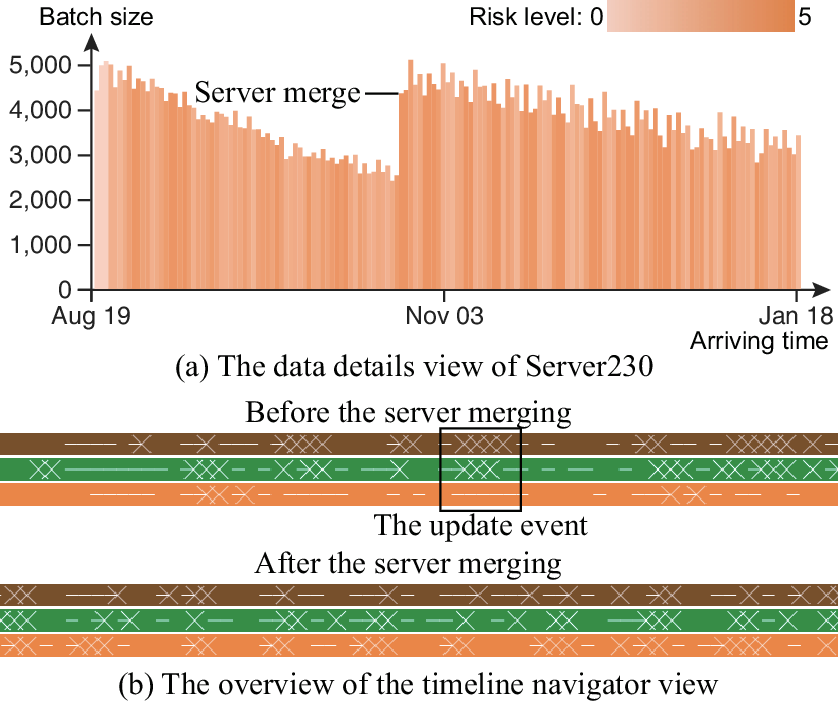}
   \caption{Visualizations of a server merging event of Sever230: (a) the data details view and (b) the timeline navigator view.}
   \label{fig:net_ove}
\end{figure}

We first browse the entire time span to learn about the evolution of consumption behaviors in three servers.
%Because the time span is relatively short, the entire training process of the prediction model is explored. 
With the threshold of $70\%$, all concept drifts are identified as inconsistent by the consistency judgment model. Besides, the projection of parameters of Server230 is far from those of the parameters of other data sources (see Figure~\ref{fig:net_wpv}, \textbf{DR2}), which implies that the consumption behaviors of the players in Sever230 is quite different from the other two servers. In particular, Server164 has a similar trajectory with Server17 from August 2013 to November 2013. %Analysts also notice that the Server230 has the lowest accuracy from beginning to end. This is because players of Sever230 (started later than other two) are more diverse, leading to complicate concepts. 
After that, the trace of Server230 shows a sharp downward turn. The specific time point is further studied. As shown in Figure~\ref{fig:net_ove}(a), the number of players at the moment was doubled---on October 24th, the game operators merged Server230 with Server229 to maintain player engagement. We notice that Server230 has fewer concept drifts than the other two servers before this merge, as shown in Figure~\ref{fig:net_ove}(b). We come up with a hypothesis that the consumption behaviors of players in Server230 affected less by various events than other servers. This phenomenon may be one reason for the operators to merge servers. 

\begin{figure} [htbp]
   \centering
   \includegraphics[width=0.99\columnwidth]{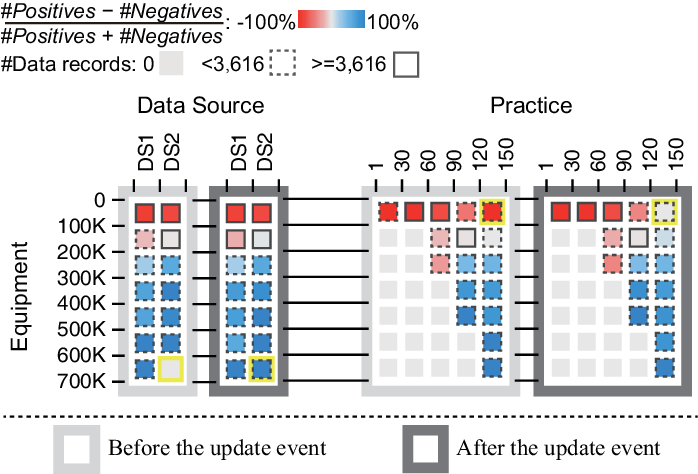}
   \caption{Patterns before and after the update event. Blue indicates that more players have consumption in the upcoming seven days.}
   \label{fig:net_act}
\end{figure}

To verify this hypothesis, an activity held a month earlier than the server merge (see Figure~\ref{fig:net_ove}(b)) is analyzed. The consistency judgment model regards Server230 as inconsistent with the other two servers. We use records from Server17 and Server164 to study players' respond to this event. The time segments before and during the event are selected separately (\textbf{DR3}). As shown on the left of Figure~\ref{fig:net_act}, the right-bottom square changes from gray to blue, which implies that a certain number of players in Server164 (DS2) took the opportunity to update their \textbf{equipment} to the highest level (\textbf{DR4, DR5}). Besides, some high-\textbf{practice} but poorly equipped players in both servers were enthusiastic about the event and consumed virtual currency during the event (see the left-top squares in the two cells on the right of Figure~\ref{fig:net_act}). However, no significant changes are observed in Server230. 

We invited a data analysis expert, who was in charge of operating the game, to check our findings. She told us that because of player loyalty, the older the server is, the more the enthusiasm for game events. For newly opened servers, payment peaks occurred mainly at the moment of launching. As for merging servers, she told us that some players created smurfs in servers to collect equipment or provide assistance after merging servers. And the most efficient way to create a high-quality smurf is to consume during events. These observations verify the hypothesis.

\begin{figure*} [!ht]
   \centering
   \includegraphics[width=1.99\columnwidth]{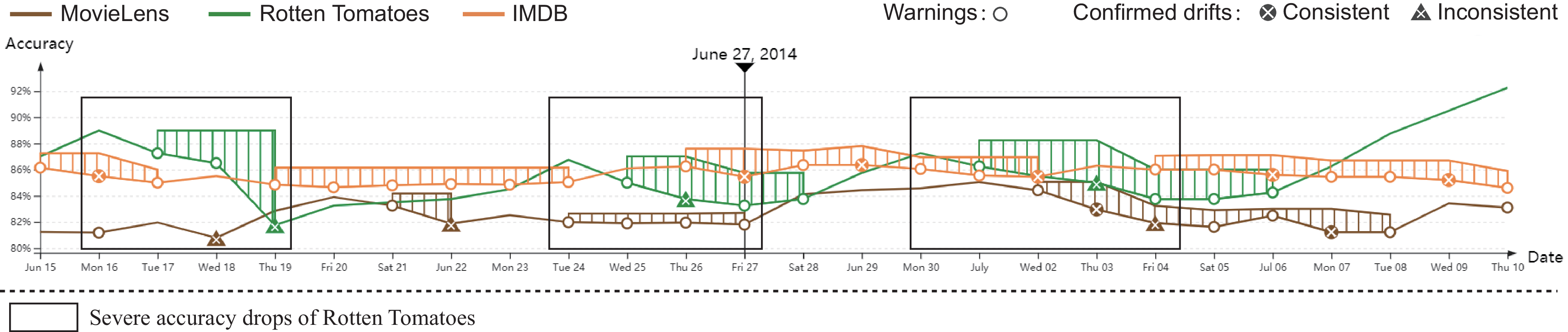}
   \caption{The accuracy fluctuation of the prediction models trained by the data from three data sources (June 15th, 2014 - July 10th, 2014).}
   \label{fig:mov_olv}
\end{figure*}

\subsection{Movie Rating Prediction}

To comprehensively learn about the evolution of audience preference on different movies, we study whether the average rating of a movie will increase in the next seven days from three platforms: Rotten Tomatoes~\cite{rottenTomato} (recorded reviews from critics), IMDB (collected from Twitter)~\cite{IMDBtweet}, and MovieLens~\cite{movielens}. By extracting data stamped in the common time segment (from February 28th, 2013 to March 31st, 2015), there are $96174$, $385015$, $1127948$ records from three sources, respectively. Each record includes rating date, rating score, and movie ID.  The movie ID is replaced with the movie description~\cite{movie}, like the release \textbf{year}, \textbf{budget}, \textbf{duration}, etc. The training data for the prediction model has $15$ dimensions. 

\begin{figure} [!htbp]
   \centering
   \includegraphics[width=0.99\columnwidth]{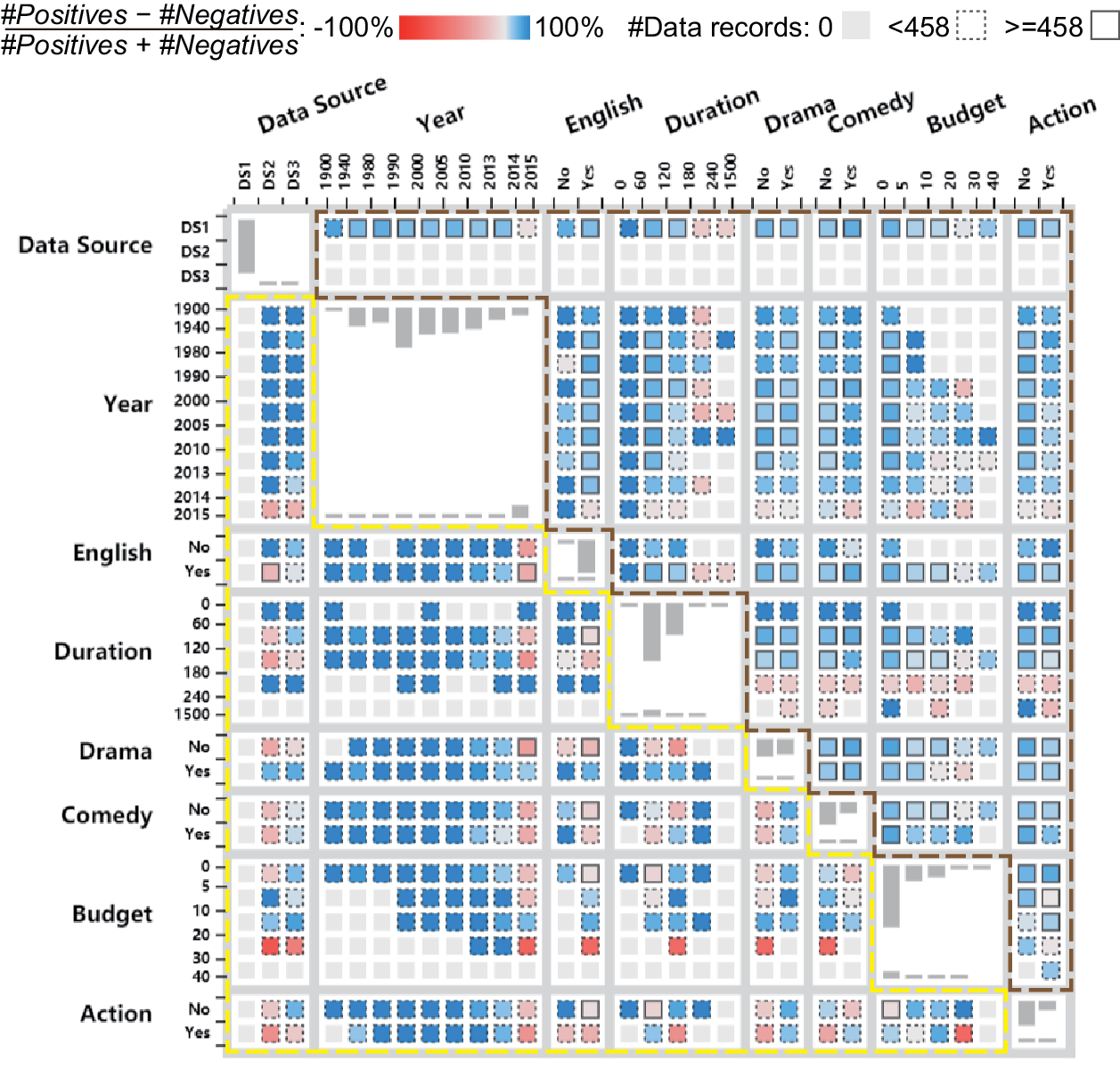}
   \caption{The correlation matrix displays concepts extracted from the segment around June 28th, 2014. The lower-left corner (the dashed region in yellow) shows records from Rotten Tomatoes (DS2) and IMDB (DS3). The upper-right corner (the dashed region in brown) shows those from MovieLens (DS1). Due to the uneven distribution of labels, the color mapping of the correlation matrix is reset to map the average difference to white.}
   \label{fig:mov_tf4}
\end{figure}

Our analysis starts from the summer vacation because most people have chances to watch movies during this period. We select the segment from June 15th, 2014 to July 10th, 2014 from the timeline navigator view. As shown in Figure~\ref{fig:mov_olv}, the prediction models trained by the data records from different data sources have distinct accuracy fluctuations (\textbf{DR2}). %Compared with Rotten Tomatoes, the fluctuations of IMDB and MovieLens are smoother. Because three data sources exhibit inconsistency, we study data records separately (\textbf{DR3}). 
The consistency judgment model suggests to study three data sources separately. After grouping tests (\textbf{DR3}), we find that the records from Rotten Tomatoes and IMDB (the lower-left corner highlighted by the yellow dashed line) show clearer patterns than those from MovieLens (the upper-right corner highlighted by the brown dashed line, \textbf{DR5}), as shown in Figure~\ref{fig:mov_tf4}. By studying red squares in the lower-left corner,  we detect three descriptions corresponding to the movies whose ratings have declined: movies whose release \textbf{year} is 2014, movies with relatively high \textbf{budget}s and \textbf{action} movies (\textbf{DR4}). Moreover, squares corresponding to each intersection of the above descriptions are in conspicuous red. The reason may be that a highly anticipated movie does not meet audience expectations. The rating records turn out that the disappointing movie is \textit{Transformers: Age of Extinction}.

\begin{figure} [htbp]
   \centering
   \includegraphics[width=0.99\columnwidth]{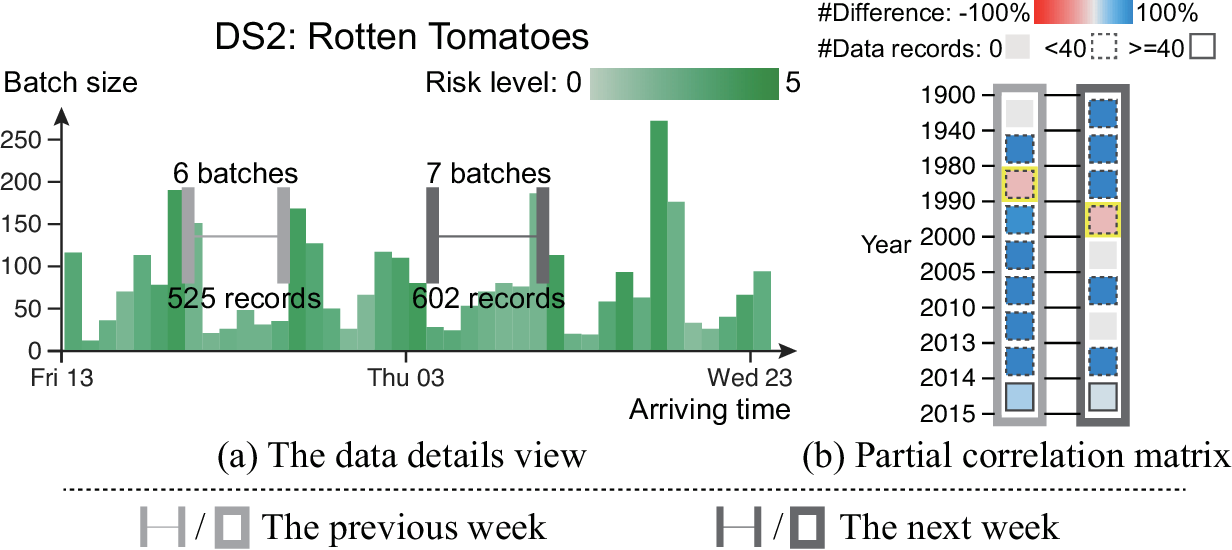}
   \caption{The details of data records collected from Rotten Tomatoes in two weeks. (a) The data details view shows two specific time segments. (b) Two cells of the correlation matrix depict the correlation between \textbf{year} and the concept.}
   \label{fig:mov_rtw}
\end{figure}

We further observing the accuracy curves to learn platform characteristics. It can be seen from Figure~\ref{fig:mov_olv} that the accuracy fluctuation of Rotten Tomatoes is more severe than others. Especially, there exists periodic fluctuations in the curve (\textbf{DR2}). Concept drifts appeared once in about a week. The number of arriving data records has the same periodicity, as shown in Figure~\ref{fig:mov_rtw}(a). We select the time segments of the previous week and the next week (\textbf{DR3}). The main difference is identified from movies released in the 1990s and 2000s (see Figure~\ref{fig:mov_rtw}(b), \textbf{DR4}, \textbf{DR5}): new ratings reduce the average ratings of certain old movies. Similar patterns are not found from other data sources. This may be caused by specific recommendations from the Rotten Tomatoes---we notice that there are sections for ``hidden gem movies'' on the Rotten Tomatoes website.

\section{Discussion}
In this section, we summarize the feedback from three experts and discuss the considerations of our approach.
\subsection{Expert Reviews}
We invited three professors in related fields as experts to review our system. The first expert (E$1$) has been working on massive data analysis for twelve years. The other two experts (E$2$ $\&$ E$3$)  have at least seven years of experience in visual analytics of time-series data.
A semi-structured interview was conducted with each expert through a remote conference. We first introduced our method and visual design in about 20 minutes. Then, we showed them case studies, during which they were free to ask questions and express opinions. We summarized their feedbacks as follows.
%After a 30-minute introduction, E$1$ showed a keen interest in our case studies and requested to visit our online system to learn more details. 
%The introduction took an hour and 40 minutes for each expert. 
%Experts asked several questions about usage details, like how the patterns in the timeline navigator view lead to further analysis. We answered these questions through three case studies mentioned in Section~\ref{sec:cas}. 

\textbf{Effectiveness.} All experts agree with the effectiveness of our approach. ``The concept drift index can indeed reflect the change process of the transformation data distribution over time to a certain extent,'' E1 commented.  E$2$ and E$3$ also appreciated our idea of applying the drift level index. E$3$ said that the index can effectively support interactive exploration of unknown concepts and concept drifts. 

\textbf{Scalability.} E1's main concern is whether high-dimensional data, \textit{i.e.}, data with hundreds of dimensions, can be applied in our system. Through system demo, we proved to him that our visual analysis approach is minimally affected by the curse of dimension. Related discussion can be found in Section~\ref{sec:sca}. In summary, he believes that our workflow and system can meet the need for analyzing time-series data and he would like to use our system when he has related analysis requirements. For further extension of our approach, he gave us two suggestions: 1) considering a multi-model hybrid prediction method to enhance the reliability of the concept drift index; 2) recommending concepts or concept drifts based on data features automatically. 

\textbf{Visual Designs.} Concerning the interface design, all experts gave positive feedbacks. E$2$ particularly likes the hierarchical abstraction in the timeline navigator view and the prediction model view. E$3$ was impressed by the shifted stripes in the prediction model view. 

\textbf{Learn costs.} E$2$ and E$3$ commented that analysts need time to learn before they can use the system. Considering that the definition and visual representation of concept drift are abstract and complex, they agree that it does worth learning costs. 

\textbf{Advice.} Considering that the analysis may only involve partial data sources, E$2$ suggested supporting the filter of data sources in the data entrance, which can facilitate analysts to focus on certain data sources. We update the system and allow analysts to control the display of data sources. However, the training of the consistency judgment model can not be completed interactively. Analysts have to reset data sources from the backend to modify the model results.

\subsection{The Navigation of the Drift Level Index}
The drift level index provides analysts with comprehensive navigation of various dynamics of concepts by connecting prediction models and visual analysis of time-series data. However, not all dynamics are identified by the drift level index. As mentioned in Section~\ref{sec:det}, the computation of the drift level index ignores the situations that the accuracy of predictive models is increasing or stable. The understanding of concepts keeps updating with iterations, even when the accuracy does not drop. The dynamics that can not be reflected from the drift level index mainly fall into two categories: improvements during learning processes and slow changes that can be caught by iterations. 

Prediction models initialize their understanding of concepts (\textit{i.e.}, parameters) at the beginning of the training process. The subsequent iterations always contribute to a rapid rise of accuracy. The same phenomena appear when the adaptive mechanics (replacing the weakest prediction model) are triggered to stop the accuracy declines caused by concept drifts. In other words, the results of these changes can be observed by inspecting the concepts following related concept drifts. In addition, concepts may evolve slowly. If prediction models can follow the changes by accumulative updates in iterations, no concept drift can be detected. To reveal the imperceptible changes, parameter evolution is monitored by the parameter projection view, which not only provides an overview of the entire learning process but also indicates the accumulative changes.

\subsection{Scalability}
\label{sec:sca}
\subsubsection{Visual Designs}
We discuss the visual scalability issue from the following aspects.

\textbf{Data records.} \techname{} assists analysts to locate appropriate contexts of concepts step by step, during which no attention needs to be paid on single data records or their attribute values. Because the dynamic features of data records distributed in different contexts are extracted by automatic approaches. The concepts corresponding to the selected contexts are summarized by the differences in the number of records with positive labels and negative ones. Analysts can contribute to qualitative conclusions based on the distribution of differences over an attribute or a pair of attributes, as mentioned in Section~\ref{sec:cas}. 

\textbf{Attributes.} Due to the limitation of display space, up to $15$ attributes are shown in the concept explanation view. To provide significant patterns with sufficient spaces, only attributes with top correlations with the label are listed. 

\textbf{Data sources.} Shifted stripes are employed to eliminate visual clutter caused by multiple data sources in the accuracy fluctuation chart. The gap width of stripes can be increased to insert more lines, \textit{i.e.}, adapt to more data sources. In addition, the color map that encodes data sources should also be adapted to the increasing number of data sources.

\subsubsection{Computation Time}
The performance of models are tested on a desktop with 16G memory and two Intel Core i7 6700 at 3.4 GHz and 3.41GHz processors (see Table~\ref{tab:time}). Data records from four data sources used in the first case are composed into an $88$-dimensional data source, named Case$1_{mixed}$. It can be seen that the size of data affects the performance of prediction model training. The 
computation time of drift level indices is not affected by the data dimension, but is related to the size of sliding windows. In this work, the size of sliding windows is determined according to the update frequency of data records.

\begin{table}[!htbp]
\caption{Average computation time (in milliseconds) of an iteration for the three stages with different combinations of attribute amount. The size of sliding window of drift level index is labeled in brackets.}
  \label{tab:time}
  \scriptsize
	\centering
  \begin{tabular}{p{2.5cm}|p{1.3cm}|p{1.5cm}|p{1.3cm}}
  \toprule
{Data source name \quad \quad \quad\quad \quad \quad(\#Attribute)}
& {Prediction model} & {Drift level index calculation (size)} & {Consistency judgment model}                                                                  \\ 
\midrule
MovieLens ($15$) & $2.532$   & $0.088$ ($1500$)   & $0.028$  \\

RottenTomatoes ($15$) & $2.698$  & $0.013$ ($100$) & $0.018$ \\

IMDB ($15$) & $2.542$   & $0.036$ ($500$)  & $0.015$\\
Guanyun ($22$) & $3.152$   & $0.009$ ($100$) & $0.034$\\
Tiantan ($22$) & $3.203$   & $0.009$ ($100$) & $0.022$\\
Case$1_{mixed}$ ($88$) & $7.681$   & $0.009$ ($100$) & $0.032$\\
  \bottomrule
  \end{tabular}
\end{table}

In summary, the time-consuming part of automatic approaches is training prediction models. In the current version of \techname, training and verifying are completed in the preprocessing stage. With the help of powerful computing clusters or cloud computing, it is possible to extend our system to process massive in real-time data. Hence, our system design and workflow have adequate scalability in terms of data records and attributes. 
% \subsection{The Visual Explanation of Concepts}
% As mentioned in Section~\ref{sec:def}, it is complicated to describe a concept. The relationship between environment variables and the target variable can be simulated in multiple ways, like various mapping functions. Our approach does not attempt to derive specific expressions. Instead, \techname assists analysts to explore concepts by showing the distribution of binary labels. The patterns in the correlation matrix can contribute to qualitative conclusions but can not represent quantitative relationships. Additional views are needed to facilitate further exploration of concepts. Besides, the second suggestion from E$1$ provides us a solution---recommending concepts through analysts-specified features and guide analysts to focus on what they look for.

\section{Conclusion}
In this paper, we propose a visual analysis approach to facilitate the exploration of concept drifts from multi-source time-series data.  Analysts are allowed to flexibly identify and compare the concepts with different contexts. The gradually progressive specification of the contexts is navigated by the model-derived drift level index and the consistency judgment model, which correspond to time segments and the set of data sources, respectively. A visual analysis system, \techname, is designed and implemented. 

The effectiveness of \techname{} is verified through three case studies with various real-world data sets. In addition, positive reviews are received from two experts on related fields. In the future, we plan to improve the concept explanation view to explain the relationship between attributes and the label in a more comprehensive way.

\acknowledgments{
This work was supported by National Natural Science Foundation of China (61772456, 61761136020, 61972122, 61872389) and Open Project Program of State Key Lab of CAD\&CG (A1903). This work has partially been supported by the FFG, Contract No. 854184: ``Pro$^2$Future is funded within the Austrian COMET Program Competence Centers for Excellent Technologies under the auspices of the Austrian Federal Ministry of Transport, Innovation and Technology, the Austrian Federal Ministry for Digital and Economic Affairs and of the Provinces of Upper Austria and Styria. COMET is managed by the Austrian Research Promotion Agency FFG.''}

\bibliographystyle{abbrv-doi}

\bibliography{template}
\end{document}